\documentclass[11pt,a4paper]{article}
\usepackage[T1]{fontenc}
\usepackage{lmodern,amsmath,amssymb,graphicx,booktabs,tabularx,array,xurl,flafter}
\usepackage[top=24mm,bottom=24mm,left=28mm,right=28mm]{geometry}
\usepackage[authoryear,round]{natbib}
\setcitestyle{aysep={}}
\usepackage{microtype,caption,xcolor}
\definecolor{inklink}{RGB}{35,65,91}
\usepackage[colorlinks=true,allcolors=inklink]{hyperref}
\newcolumntype{Y}{>{\raggedright\arraybackslash}X}
\input{glyphtounicode}
\hypersetup{pdftitle={The Elephant-Goldfish Model for software design review: what a fresh reviewer can see},pdfauthor={Sabry E. Farrag}}
\usepackage{titlesec,fancyhdr}
\definecolor{accent}{HTML}{1B4F72}
\definecolor{muted}{HTML}{667889}
\titleformat{\section}{\sffamily\Large\bfseries\color{accent}}{\thesection}{.65em}{}
\titleformat{\subsection}{\sffamily\large\bfseries\color{accent}}{\thesubsection}{.65em}{}
\titlespacing*{\section}{0pt}{15pt}{6pt}
\titlespacing*{\subsection}{0pt}{11pt}{4pt}
\newcommand{\orcidmark}[1]{\href{https://orcid.org/#1}{\raisebox{-.17ex}{\includegraphics[height=1.62ex]{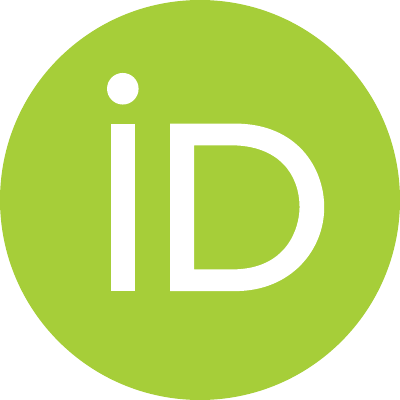}}}}
\begin{document}
\thispagestyle{plain}
\begin{center}
{\sffamily\fontsize{22}{25}\selectfont\bfseries\color{accent}The Elephant-Goldfish Model for software design review: what a fresh reviewer can see\par}
\vspace{10pt}
{\large Sabry E. Farrag\,\orcidmark{0009-0002-0735-7443}\par}
\vspace{5pt}
{\small School of Architecture, Computing and Engineering\\
University of East London, London, United Kingdom\par}
\vspace{3pt}
{\small Corresponding author: \href{mailto:sabryelgatem@gmail.com}{sabryelgatem@gmail.com}\par}
\end{center}
\vspace{3pt}
{\color{muted}\hrule}\vspace{8pt}
\noindent{\sffamily\bfseries\color{accent}Abstract}\par\nobreak\smallskip
\noindent A design can be easy to understand and still specify the wrong system. Google's Elephant-Goldfish Model asks fresh agents to review a design without the conversation that produced it. We examined how this choice affects detection of omissions, agreement violations and internal defects. We seeded nine defects in each of 47 public software design documents and supplied each with a scripted drafting conversation. A fresh reviewer given the conversation as its own earlier exchange detected 94.8\% of omitted-decision instances, compared with 62.4\% for the same model given the document alone. It detected 98.3\% of violations of conversation-only requirements. On document-visible defects, recall was 87.5\% with the conversation and 88.4\% without it, equivalent within a prespecified 15-percentage-point margin. Three model judges scored detection and were calibrated on constructed controls. In document-only runs, the plugin's five-stage review took a geometric mean of 9.03 minutes against 1.81 for one reviewer. Its estimated recall was lower under both judging procedures, with a difference distinguishable from zero under one. A later generic-prompt comparison gave a similar result. The experiment separates two objectives for automated design review: testing whether a new reader can reconstruct the design and checking whether the design preserves the decisions that govern implementation.\par
\medskip\noindent{\small\textbf{Keywords}\quad Elephant-Goldfish Model, automated design review, software specifications, coding agents, context isolation, requirements inspection\par}

\section{Introduction}\label{sec:intro}

A developer asks for pending chat commands to expire after 60 seconds. The design specifies five minutes. A reviewer can read the design carefully, explain exactly when a command expires, and endorse a design that violates the agreement. A different problem arises when the design requires event logging but drops the decision to disable disk writes during tests. Even a reader who missed the design meeting might ask how the tests isolate the filesystem. Both examples come from the same seeded document in our experiment. What a reviewer needs to know depends on which error the review is meant to catch.

The Elephant-Goldfish Model (EGM) makes this question an engineering decision. In Rensin's account, an elephant develops the design with a human over a long conversation. A goldfish reads the result in a new session and explains what it means \citep{rensin}. Google's plugin turns that idea into a workflow for drafting, inspecting and resolving software specifications \citep{google,suchy}. A fresh reader puts the document under a demanding test: can someone who missed the conversation understand what to build? A team also needs to know whether the document faithfully records what was agreed. A clear but wrong timeout passes the first test.

We study these objectives in a within-document laboratory experiment. We place defects on both sides of the boundary between conversation and specification. One defect removes an agreed decision and leaves the affected component in view. Another changes a requirement whose agreed value is available only in the conversation. A third makes the document inconsistent, incorrect or incomplete on its own terms. Every reviewer starts in a fresh session. Giving the transcript to one of those sessions changes access to the agreement while keeping model identity and session freshness fixed. The prompt also frames the transcript as the reviewer's own earlier exchange.

We ask four questions. \textbf{RQ1.} How does giving a fresh reviewer the drafting conversation change detection for each defect type? \textbf{RQ2.} What does EGM's consolidated report detect, and how long does it take, compared with one fresh reviewer? \textbf{RQ3.} How does recall transfer across two EGM configurations, and what do the reviewers detect in historical fixes? \textbf{RQ4.} Which sources of evidence do comparable tools give their reviewers? We answer the documentary question first to distinguish what the released workflow permits from what each experimental condition received.

\section{How the Elephant-Goldfish Model reviews a design}\label{sec:egm}

\subsection{From agreement to approval}
EGM assigns distinct responsibilities to the developer, the drafting session and the review sessions (Table~\ref{tab:workflow}). The elephant can ask why a constraint exists, compare alternatives and revise the draft as the developer settles trade-offs. The specification becomes the material handed to another reader. A goldfish starts without the drafting history. Its access to other evidence is a separate choice: in release 0.1.1 the orchestrator can pass a reading list to reviewers and withholds it from the comprehension test. Human judgment returns when someone decides which findings justify changing the specification \citep{google,suchy}.

\begin{table}[tb]
\caption{Review activities, inputs and outputs in EGM v0.1.1. The experiment used the review command on completed drafts.}\label{tab:workflow}
\small
\setlength{\tabcolsep}{4pt}
\renewcommand{\arraystretch}{1.12}
\begin{tabularx}{\linewidth}{@{}>{\raggedright\arraybackslash}p{.235\linewidth}YY@{}}
\toprule
\textbf{Activity} & \textbf{Evidence available} & \textbf{Output or decision}\\
\midrule
Explore and draft & Project context, requirements and discussion with the developer & Agreed concepts brief and design draft\\
\addlinespace[5pt]
Test comprehension & Target document alone, read by a fresh agent & Paraphrase of the proposal and existing system, identifying gaps\\
\addlinespace[5pt]
Inspect the design & Draft, stage instructions and any supplied reading list & Findings on design choices, technical details and implementation readiness\\
\addlinespace[5pt]
Consolidate and resolve & Stage reports, design and developer priorities & Consolidated report, decisions on findings and an approved or revised design\\
\bottomrule
\end{tabularx}
\end{table}

The five review stages are an initial high-level review, comprehension, technical critique, a second high-level review and implementation readiness. The comprehension reader first paraphrases the proposal and existing system. A technical critic can check another relationship, such as agreement between a specification and its referenced component. Consolidation merges and prioritizes observations before a developer addresses them. Our primary workflow outcome is the report that reaches the developer.

\subsection{RQ4. What can this reviewer know?}
Four sources distinguish review configurations: the drafting conversation, the proposed document, referenced code, and an independent statement of requirements. Additional evidence changes the meaning of a successful comprehension test. For a conformance check, withholding the governing requirement can make the correct answer unknowable. We coded the documented permissions of eight tools at fixed revisions. Seven describe a review step. Of these, four prescribe a fresh context, one uses the drafting discussion, and two leave conversation access unspecified. Four direct a reviewer to code and two explicitly identify original requirements as an input. These are documented permissions; run records establish actual inputs for the configurations we executed.

EGM permits referenced code. The Cross-Context Review implementation accepts the artifact alone. The cc-sdd design reviewer can use discussion history, requirements and code. A timeout conflict has a different chance of being observable in each. The evidence map in Online Resource~1 identifies all eight tools and their revisions \citep{google,songref,ccsdd,bmad,superpowers,speckit,openspec,agentos}.

\subsection{Relation to inspection and automated review}
Scenario-guided inspectors detected more requirements faults than ad hoc or checklist readers in a controlled experiment \citep{porter95}. Teams that pooled inspectors reading from different perspectives extended their coverage \citep{basili}. Collection meetings could discard findings, while pooling independent readers could raise more false alarms \citep{porter95,porter97}. EGM brings both issues into automated review: personas distribute attention, and consolidation determines what survives. We compare the delivered report with a single review and separately examine the pool of three independent reviews.

Design knowledge can remain in an architect's head after a decision has been recorded \citep{jansen}. Agent-assisted development gives that knowledge another location, the drafting conversation. Practitioners of specification-driven development describe delegating review to agents and selectively reading specifications \citep{agarwal}. Studies of architectural decisions have evaluated language models supplied with design context \citep{dhar}; document-consistency experiments have shown that even the representation of a table changes detection \citep{fukuda}. \citet{gaertner} formalized requirements to detect contradictions between statements. Our timeout example tests a different relation: a consistent specification conflicts with a decision recorded elsewhere. In issue specification, \citet{wangagr} varied the retrieval mechanism while holding the generation model and prompt fixed. Our transcript contrast fixes the model and review request and varies access to the agreement together with its authorship framing.

Cross-Context Review evaluated fresh sessions on code, tutorials and presentation scripts. Its reanalysis left the difference from one same-session review unresolved \citep{songccr,songcross}. Models can favour their own outputs \citep{panickssery}, and long contexts can make evidence in the middle harder to retrieve \citep{liu}. Both observations motivate a cold read. \citet{jin} showed that reviewers judging code against requirements can reject correct implementations and used executable checks to investigate those judgments. We examine an earlier engineering decision: which behaviour the specification should require before an implementation exists. Our literature search, repeated on 21 November 2026, found practitioner reports about EGM and no indexed controlled comparison of the plugin with a single fresh reviewer in the databases searched. Online Resource~1 records the search boundary.

\section{Study design}\label{sec:method}
We conducted a paired laboratory experiment on pre-implementation documents \citep{stol,wohlin}. Each seeded design went to all four primary review conditions. Holding the document fixed lets us compare which individual defects reach the report in each condition.

\subsection{Documents and defects}
We searched EGM's design directory, other public design directories and filenames used for EGM outputs. Of 469 candidates, 154 met the inclusion rules (15 from EGM's repository and 139 from other repositories). We held out seven pilot documents and eleven documents represented in the historical set. The sample of 48 deliberately included all eleven remaining EGM documents alongside 37 from other repositories. One drawn document failed to produce a seed list after 12 attempts. The 47 analysed documents came from 31 repositories. The search used a 100-line minimum and a character-based language filter; one mainly Portuguese document passed. Documents receive equal weight. Weighting the two strata by frame size changed every observed contrast by at most 1.6 percentage points.

A seeding agent introduced three omitted decisions, three conflicts with conversation-only requirements and three document-visible defects per document. The last class covers contradictory statements, missing failure cases and wrong facts. In the chat-command design, the timeout value requires the developer's agreement to check. A claim of ten emitted events can be checked against the twelve entries in the list. For conversation-only conflicts we rejected seeds flagged by document-only probes, including three probes using the single-reviewer procedure. Low detection without the conversation is therefore a manipulation check. Omitted decisions had no equivalent visibility filter.

A separate writer session produced both sides of each drafting conversation. Each removed decision and governing requirement appeared once. Seven or eight additional agreements in the same form concerned details the document stated or satisfied. The main transcripts contained 1,802--2,429 words. A long-transcript arm preserved the targets in 9,912--11,021 words for 15 documents. A historical arm used 11 pre-fix documents, two from EGM, each with one defect anchored to a later correcting commit. Its transcripts draw only on the pre-fix document. This arm tests detection on natural defects without introducing additional conversation evidence.

\subsection{Review conditions}
Table~\ref{tab:arms} defines the configurations. C1 and C2 used the same Opus model and the same request naming the kinds of defect sought. C1 also received the transcript, framed as its own earlier exchange. C3 ran the released plugin in Claude Code using native persona subagents. C4 ran the same skill and persona files through OpenCode's general-subagent route with MiMo configured. Both used the uninterrupted preset, an empty reading list and the target document without repository code. Provider, harness and dispatch change together between C3 and C4.

\begin{table}[tb]
\caption{Inputs and replication. Counts refer to 47 seeded documents unless otherwise indicated. Every session starts fresh.}\label{tab:arms}
\small\begin{tabularx}{\linewidth}{@{}lYrl@{}}\toprule
ID & Reviewer and material supplied & Runs/doc. & Timing\\\midrule
C1 & Single Opus reviewer, transcript and document & 3 & Main\\
C2 & Single Opus reviewer, document & 3 & Main\\
C3 & EGM, Claude Code and Opus, document & 1 & Main\\
C4 & EGM, OpenCode and MiMo, document & 1 & Main\\
C1L & As C1 with a long transcript (15 documents) & 3 & Exploratory\\
C2G & As C2 with a generic review request & 3 & Post hoc\\\bottomrule
\end{tabularx}\end{table}

Jobs started in randomized order within each stage. C1 and C2 each produced 141 reports, and C3 and C4 each produced 47. The 11 historical documents received one review per main condition. After the main results, we removed the defect list from C2's request and ran this generic version three times on each main document. We fixed its wording and analysis before those runs. It used a later Claude Code release about a week after the main comparison.

\subsection{Measurement and inference}
Our outcome is whether a review reports a particular defect. Judges from three families (Opus, GLM and Qwen) received neutral finding identifiers and randomly ordered descriptions of real and sham defects. They screened the findings and reconsidered their proposed matches in fresh calls. Unanimous screening or confirmation by two judges credited detection. A fourth family wrote positive controls, planted alone and embedded within other findings. Six sham defects per document served as negative controls. Judges saw neither the reviewer condition nor the control labels. The author briefly reviewed the model judges' assessments. The reported agreement statistics describe the three model judges. Generative models assisted with preparing the materials, producing and judging reviews, and drafting and restructuring this manuscript. Online Resource~1 describes the instruments and gives the review prompts.

Pilot string matching produced condition-dependent error. The first semantic judging procedure (v3) exceeded its 5\% sham threshold for C3 on the first 12 main documents. Revised wording (v4) clarified how closely a finding must match a defect. We developed it on that stage and fixed it before judging the other 35 documents. Across the four main arms, v4 recovered 3,362 of 3,384 planted findings and credited 26 of 2,256 shams. Fleiss' $\kappa$ on the real cells was 0.954. Agreement and recognition on controls differ from accuracy on reviewers' native findings; independent human labels would directly test that relationship \citep{wang,verga}.

We averaged detection across each document's runs and compared conditions within document. Thus 423 opportunities for a defect type in C1 represent 141 distinct seeds reviewed three times. For observed recall $p$, planted sensitivity $s$ and sham rate $f$ within condition and type, adjusted recall was $(p-f)/(s-f)$ \citep{rogan}. The all-type contrast weights types equally. A document bootstrap with 2,000 resamples re-estimated calibration and contrast together. Paired $t$ intervals describe observed differences; a document-clustered logistic generalized estimating equation provided a second implementation \citep{liang}. Adjustment assumes that control error rates transfer to reviewers' own findings. A credited sham can also reflect a correctly matched false allegation.

We prespecified a practical margin of 15 recall points and used two one-sided tests for equivalence \citep{schuirmann}. A directional effect required both observed and adjusted intervals to exclude zero on the same side. Equivalence required both equivalence intervals to fall inside the margin. These conclusions can coexist for a small, precisely estimated difference. For the three transcript contrasts we used 98.33\% effect intervals and 96.67\% equivalence intervals to account for multiplicity. The workflow and configuration contrasts use 95\% and 90\%, respectively. The design freeze was locally timestamped. We distinguish results under the two judging procedures, the held-out stage, and analyses chosen after the main results.

\section{Results}\label{sec:results}

\subsection{RQ1. Access to the conversation recovers omitted decisions}
\begin{figure}[tb]
\centering\includegraphics[width=\linewidth]{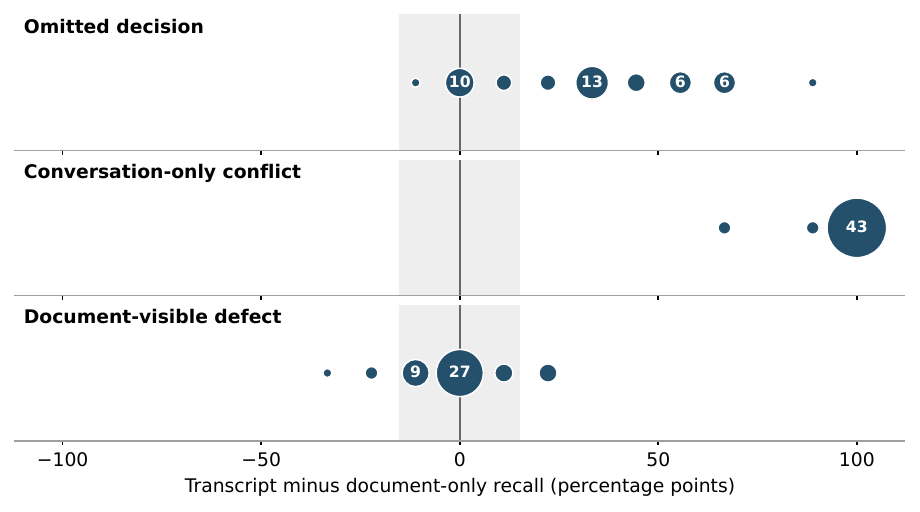}
\caption{Observed paired differences for the same 47 documents, oriented towards the transcript reviewer in all three panels. Circle area counts documents at a given difference; labels identify groups of six or more. Shading marks the prespecified 15-point margin for the mean contrast.}\label{fig:paired}
\end{figure}

Table~\ref{tab:detection} gives the observed detection counts. The reviewer given the transcript as its own exchange detected 94.8\% of omitted decisions, compared with 62.4\% for the document-only reviewer. The adjusted difference was 32.6 percentage points in favour of access (98.33\% interval 23.6 to 41.4). The direction held in 36 documents, ten tied and one favoured the document alone. Figure~\ref{fig:paired} shows the between-document distribution for all three types. The same model detected 416 of 423 conflicts with conversation-only requirements when it could read the requirement. The 1 of 423 without it verifies that the probe-based selection produced the intended visibility boundary.

\begin{table}[tb]
\caption{Detected seed--review opportunities by defect type under v4. Each short-transcript or single-reviewer row covers three runs per document; each EGM row covers one. Parentheses give observed percentages.}\label{tab:detection}
\small\begin{tabularx}{\linewidth}{@{}Yrrr@{}}\toprule
Reviewer & Omitted decision & Agreement conflict & Document-visible\\\midrule
Transcript (C1) & 401/423 (94.8) & 416/423 (98.3) & 370/423 (87.5)\\
Single (C2) & 264/423 (62.4) & 1/423 (0.2) & 374/423 (88.4)\\
EGM--Opus (C3) & 76/141 (53.9) & 2/141 (1.4) & 112/141 (79.4)\\
EGM--MiMo (C4) & 77/141 (54.6) & 6/141 (4.3) & 117/141 (83.0)\\
Generic (C2G) & 266/423 (62.9) & 1/423 (0.2) & 357/423 (84.4)\\\bottomrule
\end{tabularx}\end{table}

On defects both readers could inspect directly, C2 detected four more opportunities than C1 across 423. The adjusted difference was 0.6 points in favour of C2. The union of the two equivalence intervals spanned $-2.9$ to $+4.5$ points, comfortably inside the prespecified margin. The two conditions tied on 27 of 47 documents. A transcript thus supplied substantial missing evidence on one class of defect at little measurable cost on the document-visible class.

We examined which omissions contributed to the difference. Of the 141 removed decisions, C2 found 79 in every run, 19 intermittently and 43 in none. C1 found 110 of the 129 opportunities for the last group. A split selected by C2's misses necessarily makes its performance look poor in that group. We therefore rotated a held-out run in a descriptive reanalysis: for each omission, select it when two C2 runs missed it, then score the third C2 run and the corresponding C1 run. Across 140 eligible held-out opportunities C2 recovered 11 and C1 recovered 121. An omission can enter more than one fold, so these are descriptive counts. On the complementary 283 opportunities the counts were 253 and 280. EGM also found some omissions that C2 missed three times (9 of 43 in C3, 7 in C4). Some omitted decisions leave clues a different reading can recover. Online Resource~2 reproduces the split from individual decisions.

The transcript also helped when the agreements occupied a smaller share of the text. For the 15 documents assigned to that arm, long-transcript recall on omissions was 124/135, against 133/135 on conversation-only conflicts and 121/135 on document-visible defects. For omissions, long minus short transcript recall was $-0.044$ (95\% interval $-0.113$ to $0.025$); long transcript minus document alone was $+0.244$. The comparison concerns explicit decisions stated once and preserved amid additional material.

\subsection{RQ2. Five review stages against one fresh review}
Table~\ref{tab:contrast} shows the five planned contrasts. In the document-only setting, C3 detected 190/423 seeded opportunities (44.9\%). C2 detected 639/1,269 (50.4\%). Under v4, the adjusted difference was $-7.8$ points and the observed difference $-5.4$ points. Both intervals excluded zero and both equivalence intervals fell inside the 15-point margin. The older judging procedure also put C3 below C2, although its interval included zero. A large share of each condition's denominator consists of conversation-only conflicts that both seldom detected. Restricting the comparison to omitted decisions and visible defects produced an adjusted difference of $-11.2$ points under v4 (95\% interval $-17.7$ to $-5.3$). Equivalence within the margin then ceased to hold. All-type recall answers a question about this particular mix of defects.

\begin{table}[tb]
\caption{Paired differences in recall, in percentage points. Brackets give effect intervals at 98.33\% for the first three rows and 95\% for the last two. The adjusted estimate calibrates each condition and type separately. The v3 column is a sensitivity analysis.}\label{tab:contrast}
\small\begin{tabularx}{\linewidth}{@{}Yrrr@{}}\toprule
Contrast & Observed (v4) & Adjusted (v4) & Adjusted (v3)\\\midrule
Omission, C2$-$C1 & $-32.4$ & $-32.6$ [$-41.4,-23.6$] & $-31.2$\\
Conflict, C1$-$C2 & $+98.1$ & $+98.8$ [$+96.2,+100.7$] & $+98.0$\\
Visible, C2$-$C1 & $+0.9$ & $+0.6$ [$-3.3,+4.7$] & $+1.8$\\
All, C3$-$C2 & $-5.4$ & $-7.8$ [$-12.5,-3.3$] & $-3.9$\\
All, C4$-$C3 & $+2.4$ & $+3.8$ [$-0.6,+8.4$] & $+0.8$\\\bottomrule
\end{tabularx}
\par\smallskip\footnotesize The estimator and bootstrap are unconstrained; the conflict interval is printed before clipping at the physical ceiling.\end{table}

Two checks help interpret the workflow comparison. On the 35 documents held out from v4 wording development, C3 minus C2 was $-11.1$ adjusted points (95\% interval $-16.1$ to $-6.5$), against $+1.9$ on the first 12 ($-5.3$ to $+8.6$). Harness versions changed between stages. The stage contrast persists under v3 and therefore cannot be attributed solely to the new judge wording. Second, the later generic request (C2G) removed C2's explicit list of defects. Against it, C3 was lower by 6.4 adjusted points under v4 (95\% interval 2.3 to 10.8 lower) and by 1.5 under v3 (interval 5.7 lower to 2.5 higher). A generic single reviewer remained competitive. C2G ran later and C3 was not rerun alongside it.

Machine time gives the size of the operational trade-off. Geometric mean elapsed time was 9.03 minutes for EGM--Opus and 1.81 for the directed single reviewer, a ratio of 5.0 (95\% interval 4.66--5.32). Relative to the generic reviewer the ratio was 6.1. The consolidated reports averaged 21.3 findings against C2's 23.0. Of these, 17.72 and 18.46 respectively matched no seed. Such findings may identify genuine problems outside the seeded set. Their counts measure reading load; their correctness and severity require separate judgments.

The prespecified union of three independent C2 runs reached 54.4\% recall against C3's 44.9\%. Their combined arithmetic mean time was 5.79 minutes per document, against 9.42 for C3. Pooling exposed a reader to about 69 findings before deduplication. Conversely, pooling EGM's stage reports with its final report raised C3's observed recall to 52.0\%. Sham credits also rose, from 11/282 on final reports to 24/282 on pooled reports. The longer list creates additional opportunities for a judge to match a seed. Table~\ref{tab:overlap} shows an exploratory comparison of which defects reached each output. The EGM report added 11 detections beyond the three-review pool, nine of them omitted decisions. The pool added 51 beyond EGM, including 19 document-visible defects. Comparing EGM with each C2 run separately gave 14--18 detections unique to EGM and 36--45 unique to C2. These comparisons count distinct credited defects. The single-run comparison gives each method one delivered report per document, with the machine times reported above.

\begin{table}[tb]
\caption{Overlap between the union of three directed single reviews (C2) and one consolidated EGM report (C3) per document. Counts use the primary judging procedure (v4). Each seeded defect appears in one cell.}\label{tab:overlap}
\small\begin{tabularx}{\linewidth}{@{}Yrrrr@{}}\toprule
Defect type & Both & C2 only & EGM only & Neither\\\midrule
Omitted decision & 67 & 31 & 9 & 34\\
Conversation-only conflict & 0 & 1 & 2 & 138\\
Document-visible defect & 112 & 19 & 0 & 10\\\midrule
All 423 defects & 179 & 51 & 11 & 182\\\bottomrule
\end{tabularx}\end{table}

\subsection{What reaches the developer? A trace through the released tool}
Each of the 47 analysed documents produced a comprehension report and a consolidated report under each EGM configuration: 94 paired outputs. We inspected one pair to illustrate how those files answer a question the recall score cannot. We used the first EGM-owned document in lexical order of the analysed identifiers and its native-plugin run, a reproducible post hoc selection. The design proposes support for three coding-agent hosts. The comprehension reviewer correctly recovers the three hosts and the plan to test installation, workflow and isolation separately. It then distinguishes an inference about existing hosts from a fact it can establish, and asks where the support table is located. The final report groups that observation with several terminology issues. Action A19 asks the developer to identify the file containing the support table. The observation survives consolidation in a form someone can act on.

The seed metric asks whether a report names a known defect. The paraphrase tests whether the proposal can be reconstructed; the action list shows which uncertainty the workflow carries forward. These are three different records of the same review. This trace establishes one observed path through the pipeline. Online Resource~1 gives the anchors and the selection rule; the archive indexes stage and final files for both configurations.

\subsection{RQ3. Configuration transfer and historical fixes}
EGM in OpenCode with MiMo detected 200/423 opportunities (47.3\%). Its adjusted advantage over EGM--Opus was 3.8 points, with the joint equivalence region from $-1.1$ to $+7.6$ points. This result concerns recall in the two configurations. Their operational costs differed sharply. The OpenCode run had a geometric mean elapsed time of 59.65 minutes per review (6.6 times the Opus configuration) and returned 9.1 more unmatched findings on average (95\% interval 6.81--11.36). The comparison cannot separate the contributions of model, harness and subagent dispatch.

On the 11 historical corrections, C1, C2, C3 and C4 flagged 3, 4, 3 and 5 respectively. A correcting commit identifies one real issue but leaves other legitimate findings unlabelled. With one review per condition and overlapping intervals, these counts describe coverage of individual cases. The seeded and historical sets answer different questions about the instruments.

\section{Implications for automated design review}
\subsection{Choose the input from the question}
A first reading without the drafting history and an agreement check serve different engineering purposes. A team deciding whether to hand a draft to a new implementer can ask a fresh reader to reconstruct it and identify its guesses. A team deciding whether the draft preserves a negotiated timeout needs access to the timeout decision. Conflating the checks can make a fluent summary look like evidence of conformance. The reviewer supplied with the full scripted conversation performed well on the second task under the framing we tested. A shorter decision record could make the agreement available at lower reading cost. Its fidelity, especially after requirements change, is an empirical question.

The distinction also gives tool builders a useful interface choice. Label a finding by the material needed to verify it: two passages in the draft, a decision in the conversation, or a specification claim and a referenced file. That label would let a developer choose the evidence to inspect before changing the design. The third category is especially important for EGM. The plugin supports referenced sources and assigns reviewers to check fit with code. Our document-only run provides a baseline for a crossed experiment in which EGM and a single reviewer each inspect the same designs with and without frozen source snapshots. Defects whose truth requires the code would give that comparison a discriminating endpoint.

\subsection{Evaluate the deliverable at the right stage}
For a developer who receives a consolidated report, a defect found early contributes only if the actionable information reaches that report. For a developer asking whether a specification stands alone, the paraphrase can itself be valuable, even when it yields no seed match. A benchmark can preserve both outcomes by storing the paraphrase, each finding's evidence and its disposition. The existing file trail makes this feasible. The next practical measure is whether an engineer accepts the finding and whether the ensuing edit corrects the design; report length alone cannot answer either question.

\section{Threats to validity}
\textbf{Sampling and experimental conditions.} Comparing each document with itself controls for differences in document difficulty. The conversation writer stated each agreement explicitly; real sessions may reverse decisions or embed them in tool outputs. The transcript reviewer did not author the design and was prompted to treat the exchange as its own. A comparison with a genuinely continuing author session would measure the additional effect of authorship and session history. Eleven documents came from EGM's own repository. A post hoc repository bootstrap produced observed 95\% intervals of $-39.4$ to $-23.7$ points for C2 minus C1 on omissions and $-10.3$ to $-1.8$ for C3 minus C2 overall.

\textbf{Practical significance.} The 15-point tolerance represents a design choice. One defect in a safety-critical or high-cost decision may matter more than several defects in exposition. The visible-defect comparison supported a tighter margin of approximately 4.5 points, while the full C3--C2 comparison needed about 11.8 points. A team can read those bounds against the cost of a miss in its own setting.

\textbf{Measurement and transfer.} The same model family prepared materials and supplied three of the four primary reviewers. The two outside judges and constructed controls partly address common-model bias. The v4 and v3 results bound the interpretation of the small workflow difference: on 114 real or sham cells where procedures disagreed, two additional model adjudicators split on 81. The large omission contrast and equivalence on visible defects held under either procedure. Inferences about EGM apply to one release in unattended review, with no repository files and no human resolution step. A single EGM run per document and configuration leaves the repeatability of its exclusive detections open. The experiment measured machine minutes, detection and unmatched findings. The cost of developers' time and the quality of the implemented system remain open outcomes.

\section{Conclusion}
In the Elephant-Goldfish Model, starting a new session and deciding what that session may read are separate engineering choices. Across 47 designs, the fresh reviewer given the scripted conversation as its own exchange recovered roughly one additional omitted decision per document across the three seeded omissions, while recall on document-visible defects remained close. In a document-only setting, EGM's multi-stage report took five times the machine time of one directed fresh review and showed no recall advantage under either judging procedure. Teams evaluating a design can use a cold read to test what the document communicates, and a review against recorded decisions to test what the document promises. Each verdict then has a clear meaning for the person who must approve the design.

\section*{Supplementary information}
Online Resource~1 gives protocols, scoring details, sensitivity analyses, the read-permission map and the worked trace. Online Resource~2 provides machine-readable detection decisions, elapsed times, calibration and contrast summaries, and scripts that reproduce the counts, defect-level overlap, held-out omission analysis and intervals for all five planned contrasts.

\section*{Statements and Declarations}
\textbf{Data availability.} The analysis data supporting the reported comparisons accompany this article as Online Resource~2. It includes the scripted transcripts, single-reviewer reports, 94 pairs of EGM comprehension and consolidated reports, source revisions, hash-verified transformations for all 47 seeded designs, detection labels under both judging procedures and elapsed times. Replaying the analyses is independent of model access. Fresh review runs require the corresponding hosted models. Online Resource~1 identifies the review and scoring instructions.

\textbf{Funding and competing interests.} The author reports no funding and no competing interests.

\textbf{Author contribution.} Sabry E. Farrag conceived and designed the study, prepared the materials, conducted the experiments and analysed the results.

\textbf{Ethics.} The study used public software design documents and model-generated reviews and involved no human participants.

\setlength{\bibsep}{3pt plus 1pt}

\end{document}